\documentclass[prl,showpacs,twocolumn]{revtex4}

\newcommand {\etal}{{\it et al.}}

\newcommand {\CFO}{CuFeO$_2$}

\newcommand {\CFGO}{CuFe$_{1-x}$Ga$_{x}$O$_2$}

\newcommand {\Spini}{\vec {S}_i}
\newcommand {\Spinj}{\vec {S}_j}

\newcommand{\Ew}{E^\omega}
\newcommand{\Hw}{H^\omega}

\usepackage{graphicx}
\usepackage{bm}
\usepackage{pifont}
\usepackage{amsmath}

\begin{document}

\title{Electromagnons in the spin collinear state of a triangular lattice antiferromagnet}

\author{S. Seki$^1$, N. Kida$^{2}$\onlinecite{Aff}, S. Kumakura$^1$, R. Shimano$^{2,3}$, and Y. Tokura$^{1,2,4}$} 
\affiliation{$^1$ Department of Applied Physics, University of Tokyo, Tokyo 113-8656, Japan \\ $^2$  Multiferroics Project, ERATO, Japan Science and Technology Agency (JST), Tokyo 113-8656, Japan \\ $^3$ Department of Physics, University of Tokyo, Tokyo 113-0033, Japan \\ $^4$ Cross-Correlated Materials Research Group (CMRG) and Correlated Electron Research Group (CERG), RIKEN Advanced Science Institute, Wako 351-0198, Japan}

\date{August 26, 2010}

\begin{abstract}
Terahertz time-domain spectroscopy was performed to directly probe the low-energy (1-5 meV) electrodynamics of triangular lattice antiferromagnets CuFe$_{1-x}$Ga$_x$O$_2$ ($x= $ 0.00, 0.01, and 0.035). We discovered an electromagnon (electric-field-active magnon) excitation at 2.3 meV in the paraelectric $\uparrow \uparrow \downarrow \downarrow$ collinear magnetic phase, while this electromagnon vanishes in the ferroelectric helimagnetic phase. Anti-correlation with noncollinear magnetism excludes the exchange-striction mechanism as the origin of dynamical magnetoelectric coupling, and hence evidences the observation of spin-orbit coupling mediated electromagnon in the present compound.
\end{abstract}
\maketitle

Magnetoelectric (ME) effect, i.e. electric (magnetic) control of magnetic (dielectric) properties, has long been an important issue in condensed matter physics\cite{Review}. While magnetic and dielectric properties usually show minimal coupling, recent discoveries of magnetically-induced ferroelectricity in frustrated magnets have enabled unprecedentedly large and versatile ME responses\cite{Kimura, TbMn2O5, Ishiwata}. Jia {\etal} suggested that at least three microscopic ME coupling mechanisms can be considered: When a ligand ion is placed at the center of two magnetic ions, magnetically-induced local electric polarization $\vec{P}_{ij}$ is described as
\begin{eqnarray}
\vec{P}_{ij} = & \vec{\Pi} (\Spini \cdot \Spinj) + A \vec{e}_{ij} \times (\Spini \times \Spinj) \nonumber \\
& + B [(\vec{e}_{ij} \cdot \Spini)\Spini - (\vec{e}_{ij} \cdot \Spinj)\Spinj],
\end{eqnarray}
where $A$ and $B$ are coupling coefficients, $\vec{\Pi}$ is a vector unique to the underlying crystal structure, and $\vec{e}_{ij}$ an unit vector connecting two magnetic moments ${\Spini}$ and ${\Spinj}$, respectively\cite{Jia1}. The first term represents the exchange striction, and is considered as the origin of ferroelectricity in some collinear antiferromagnets like Ca$_3$CoMnO$_6$\cite{Choi}. The second term comes from the inverse effect of Dzyaloshinskii-Moriya (DM) interaction\cite{Katsura}, which well explains the ferroelectricity in many noncollinear spiral magnets such as $R$MnO$_3$\cite{Kimura, DyMnO3} and Ni$_3$V$_2$O$_8$\cite{Ni3V2O8}. The third term arises from spin-dependent modulation of covalency (hybridization) between metal $d$-state and ligand $p$-state\cite{Jia1}, while this term usually oscillates and cancels out within the crystal. Note that the second and third terms in Eq. (1) rely on the spin-orbit interaction, but the first term (exchange striction) does not\cite{Jia1}. With any mechanism, a modification of magnetic structure, e.g. by external magnetic field $H$, leads to a significant change of induced electric polarization $P$.

One important consequence of such a strong ME coupling is the appearance of a novel collective excitation called electromagnon (i.e. magnon driven by a.c. electric field $\Ew$) in the dynamical regime. With detailed polarization analyses of absorption spectra, existence of electromagnon (EM) excitation has been established for ferroelectric (FE) helimagnets $R$MnO$_3$\cite{EM_First, EM_DyMnO3}, $R$Mn$_2$O$_5$\cite{EM_RMn2O5_Exp}, and Ba$_2$Mg$_2$Fe$_{12}$O$_{22}$\cite{EM_HexaFerrite}. Here, the most crucial is the microscopic origin of dynamical ME coupling, which is not necessarily identical to that of the magnetically-induced static $P$ in the same compound. According to the inverse DM scheme, the EM excitation in FE helimagnets emerges as the rotational oscillation of spin-spiral plane and associated $P$-vector\cite{EM_Katsura}. While this rotational mode should be active only with $\Ew$ perpendicular to the spin-spiral plane, the observed selection rule for $R$Mn$_2$O$_5$ contradicted with this prediction\cite{EM_RMn2O5_Exp, EM_RMn2O5_Theory}. For $R$MnO$_3$ and Ba$_2$Mg$_2$Fe$_{12}$O$_{22}$, the selection rules remain unchanged even after the spin-flop transition under applied  $H$\cite{EM_DyMnO3, EM_HexaFerrite}. Latest theoretical studies suggested that the exchange striction mechanism can also host the EM activity in noncollinear magnets, but with the selection rule tied to the chemical lattice\cite{EM_Striction}. Importantly, this exchange-striction-induced EM is inactive in the collinear magnetic phase, since the differential polarization $\delta P_{ij} \propto S_i \cdot \delta S_j$ becomes always zero ($S_i \perp \delta S_j$). This latter model well reproduces the observed selection rules or absorption spectra of EM in all the three helimagnetic compounds\cite{EM_HexaFerrite, EM_RMn2O5_Theory, EM_Striction, EM_Mochizuki}, whereas the firm experimental evidence of spin-orbit coupling mediated EM excitation is still lacking.

In this study, we report the experimental discovery of electromagnon excitation in the paraelectric collinear magnetic ($\uparrow \uparrow \downarrow \downarrow$) phase of triangular lattice antiferromagnets {\CFGO}. This EM mode was found to rather vanish in the FE helimagnetic phase. The anti-correlation between the electromagnon and the noncollinear magnetism excludes the exchange-striction mechanism as the origin of dynamical ME coupling, and hence suggests that the observed electromagnon is electrically activated by the spin-orbit coupling.

{\CFGO} crystalizes into a delafossite structure with centrosymmetric space group $R\bar{3}m$(Fig. 1(a)), which consists of the stacking of triangular lattices along the $c$-axis\cite{CuFeO2_FloatingZone}. Each Fe$^{3+}$ ($S=5/2$) ion is surrounded by O$^{2-}$ octahedra, and the magnetic frustration leads to several complex spin structures with propagation vector $\vec{k} = (q, q, 3/2)$\cite{CuFeO2_PhaseDiagram}. For simplicity, we define the $q$-vector ($\vec{q}$) as the in-plane component of magnetic propagation vector. {\CFO} is characterized by the collinear $\uparrow \uparrow \downarrow \downarrow$ (CM4) magnetic ground state below 11K, with commensurate $q=0.25$ and spin direction along the $c$-axis(Fig. 1(e))\cite{CuFeO2_PD}. In Ga-doped specimen with $x > 0.02$, the CM4 phase is replaced by ferroelectric proper screw (NC) magnetic phase with incommensurate $q \sim 0.202$\cite{CuFeO2_Polarization, CuFeO2_Ga1, CuFeO2_Seki2}, where spin rotates within a plane perpendicular to the $q$-vector and $P$ appears parallel to $\vec{q} \parallel [110]$ (Fig. 1(d))\cite{CuFeO2_ProperScrew}. The ferroelectricity in the NC phase can be explained by neither exchange striction nor inverse D-M mechanism. Instead, Arima suggested that the third term in Eq. (1) can induce finite $\vec{P} \parallel \vec{q}$ component on the delafossite lattice with proper screw magnetic order\cite{CuFeO2_Arima}. This model predicts that the reversal of $P$-vector is coupled with the reversal of vector spin chirality (i.e. clockwise or counter-clockwise manner of spin rotation), which was later confirmed by the polarized neutron scattering experiments\cite{CuFeO2_PolarizedNeutron}. The $x$ - $T$ phase diagram\cite{CuFeO2_Ga1} as well as the $H$ - $T$ phase diagram for the $x=0.01$ specimen (by the present study) are summarized in Figs. 1(b) and (c). Note that the transition from CM4 into NC(FE) can also be induced by $H$ applied along the $c$-axis\cite{CuFeO2_PhaseDiagram, CuFeO2_Polarization}. Starting from either of these magnetic ground states, increase of temperature ($T$) first induces partially disordered collinear (ICM1 or ICM2) magnetic phases\cite{CuFeO2_PD, CuFeO2_OPD} and then produces a paramagnetic (PM) phase. All magnetic phases other than NC(FE) are paraelectric\cite{CuFeO2_Polarization, CuFeO2_Ga1}.

Single crystals of CuFe$_{1-x}$Ga$_x$O$_2$ ($x= $ 0.00, 0.01, and 0.035) were grown by a floating zone method\cite{CuFeO2_FloatingZone}. Magnetization was measured with a SQUID magnetometer. Complex transmittance $t$ is obtained by terahertz time-domain spectroscopy, and further converted into complex refractive index $n=\sqrt{\epsilon \mu}$ using the relationship
\begin{equation}
t = \frac{2\mu}{n + \mu} \frac{2n}{n + \mu} \exp \Big [ -i \frac{\omega}{c} d(n-1)\Big ],
\end{equation}
where $\epsilon$, $\mu$, $d$, $\omega$, and $c$ represent the complex dielectric constant, complex magnetic permeability,  sample thickness, frequency of light, and velocity of light, respectively. To numerically solve Eq. (2), we make the pre-exponential factor approximate $4n / (n+1)^2$ by assuming $\mu \simeq 1$ unless otherwise noted. The detail of THz time-domain spectroscopy is described in Appendix and Ref. \cite{EM_DyMnO3}.

\begin{figure}
\begin{center}

\includegraphics*[width=8cm]{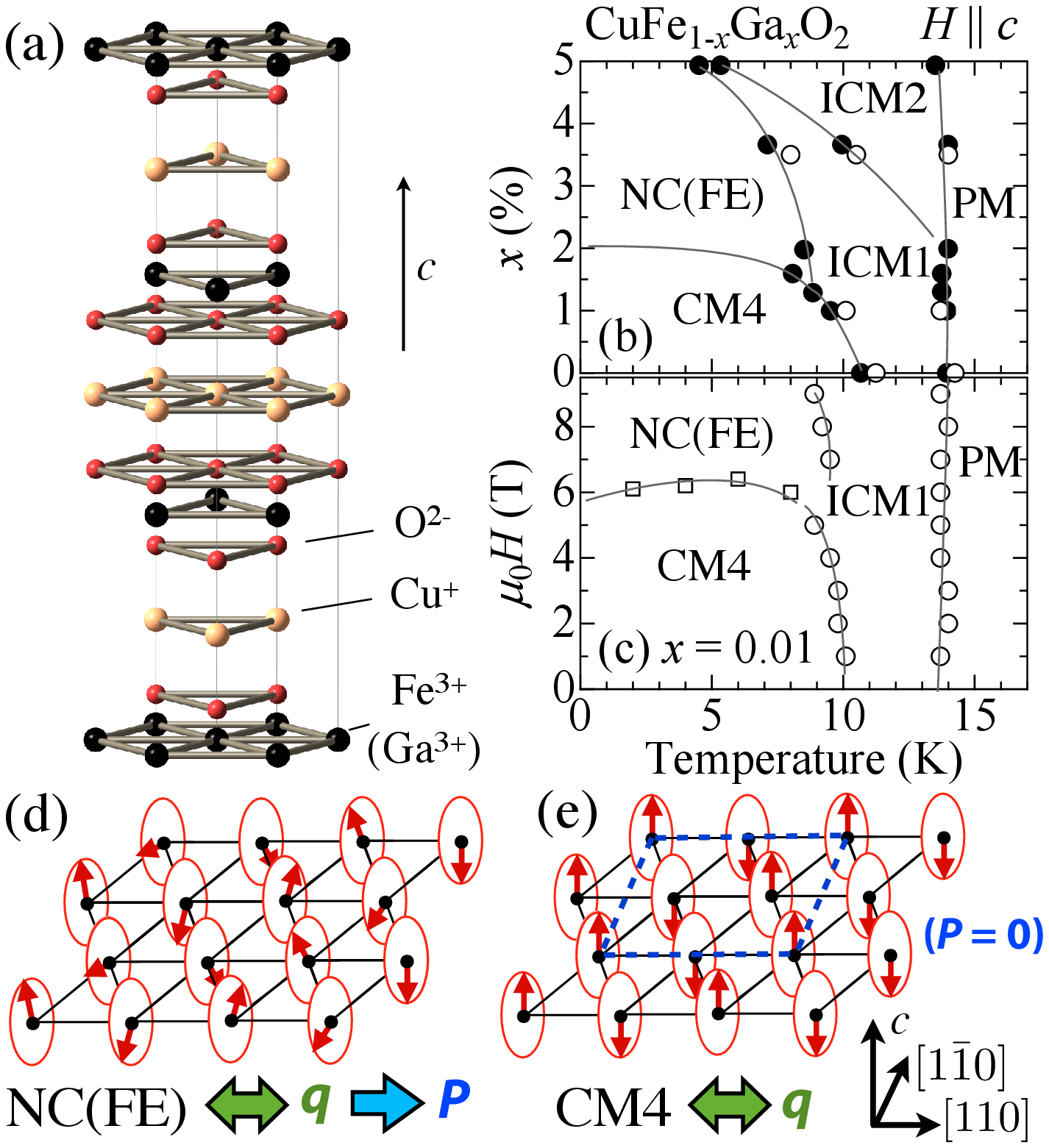}
\caption{(color online). (a) Crystal structure and (b) $x$ - $T$ magnetic phase diagram at $H=0$ for {\CFGO}. (c) $H$ - $T$ magnetic phase diagram for the $x=0.01$ specimen with static $H$ applied parallel to the $c$-axis. Circles and squares are the data points obtained from the measurements of magnetization with $T$- and $H$-increasing runs, respectively. Open (closed) symbols represent the data points determined in the present work (the previous work by Terada {\etal}\cite{CuFeO2_Ga1}). (d) and (e) indicate magnetic structures of the NC(FE) and CM4 phase, respectively. Dashed square in (e) represents the magnetic unit cell in the CM4 phase. The directions of magnetic $q$-vector and electric polarization $P$ are also indicated.}
\end{center}
\end{figure}

We first investigated the low-energy electrodynamics in the paraelectric CM4 collinear magnetic phase. Note that symmetry of triangular lattice allows the existence of three equivalent $\vec{q} \parallel \langle 110 \rangle$. The spectra observed in the present study reflect the contributions from all the three $q$-domains. Figures 2(a) and (b) indicate the real and imaginary part of $\epsilon \mu$ spectra (Re[$\epsilon \mu$] and Im[$\epsilon \mu$]) with various polarization configurations for the $x=0.01$ specimen at 4.4 K, respectively. With $\Ew \parallel [110]$ and $\Hw \parallel [1\bar{1}0]$, two resonance modes are observed at 1.2 and 2.3 meV. Only the former one survives for $\Ew \parallel [001]$ and $\Hw \parallel [1\bar{1}0]$, whereas only the latter one does for $\Ew \parallel [110]$ and $\Hw \parallel [001]$. These results unveil that the excitation at 2.3 meV is an EM mode driven by $\Ew \parallel [110]$, while the one at 1.2 meV is a conventional magnon mode driven by $\Hw \parallel [1\bar{1}0]$. 

In the following, we focus on the behavior with the $\Ew \parallel [110]$ and $\Hw \parallel [1\bar{1}0]$ configuration. To further analyze the aforementioned $\epsilon \mu$ spectrum, the corresponding absorption coefficient $\alpha ( =-2(\ln |t|)/d) $ and the decomposed $\epsilon$ and $\mu$ spectra are plotted in Figs. 3 (a)-(c), respectively. To discriminate the $\epsilon$- and $\mu$-contributions to the $\epsilon \mu$ spectrum, we first assumed $\mu = 1$ for $\hbar \omega > 2.0$ meV. The obtained $\epsilon$ spectrum can be fitted well with the sum of two Lorentzian functions; a higher-frequency mode represents the lowest-lying optical phonon to give rise to the tail absorption observed below 5 meV. By substituting this $\epsilon$-fitting function into Eq. (2), the $\mu$ spectrum is deduced for $\hbar \omega < 2.0$ meV. It can be fitted as well with a single Lorentzian function. 

\begin{figure}
\begin{center}
\includegraphics*[width=8cm]{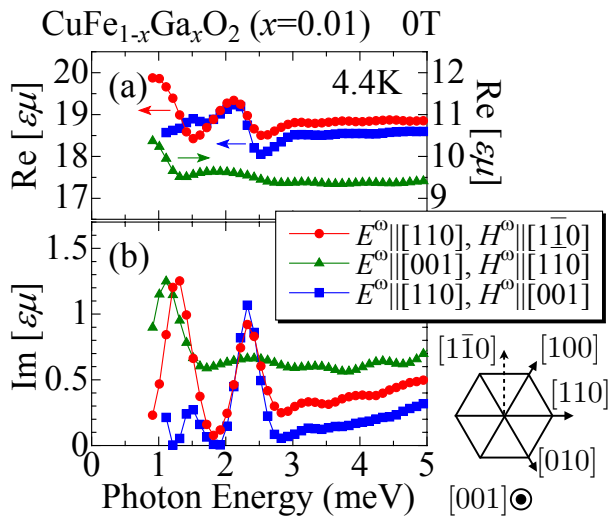}
\caption{(color online). Real and imaginary parts of $\epsilon \mu$ spectra (Re [$\epsilon \mu$] and Im [$\epsilon \mu$]) for the $x=0.01$ specimen measured at 4.4 K with various light-polarization configurations.}

\end{center}
\end{figure}

In Fig. 4(b), we show $T$-dependence of Im[$\epsilon \mu$] spectrum for the $x=0.01$ specimen. Increase of temperature leads to broadening of two resonance peaks, and they become almost undiscernible in the ICM1 magnetic phase above 10 K. ICM1 and ICM2 are partially disordered magnetic phases\cite{CuFeO2_PD, CuFeO2_OPD}, and may lose the spin correlation enough for magnons or EM excitations to be observed. The undoped $x = 0.00$ specimen also has two resonance modes at the same frequency in the CM4 magnetic ground state (Fig. 4(a)), and shows similar $T$-dependence of Im[$\epsilon \mu$] spectrum as observed for the $x = 0.01$ specimen. In contrast, the $x = 0.035$ specimen with the NC(FE) ground state shows no discernible peak structure in the whole temperature range (Fig. 4(c)). We also measured the Im[$\epsilon \mu$] spectrum for the $x=0.035$ specimen with $E \parallel [001]$ and $H \parallel [1\bar{1}0]$, but no peak structure was discerned. To summarize, the EM excitation driven by $\Ew \parallel [110]$ is active only in the paraelectric collinear CM4 magnetic phase, not in the ferroelectric NC helimagnetic phase. 

\begin{figure}
\begin{center}
\includegraphics*[width=8cm]{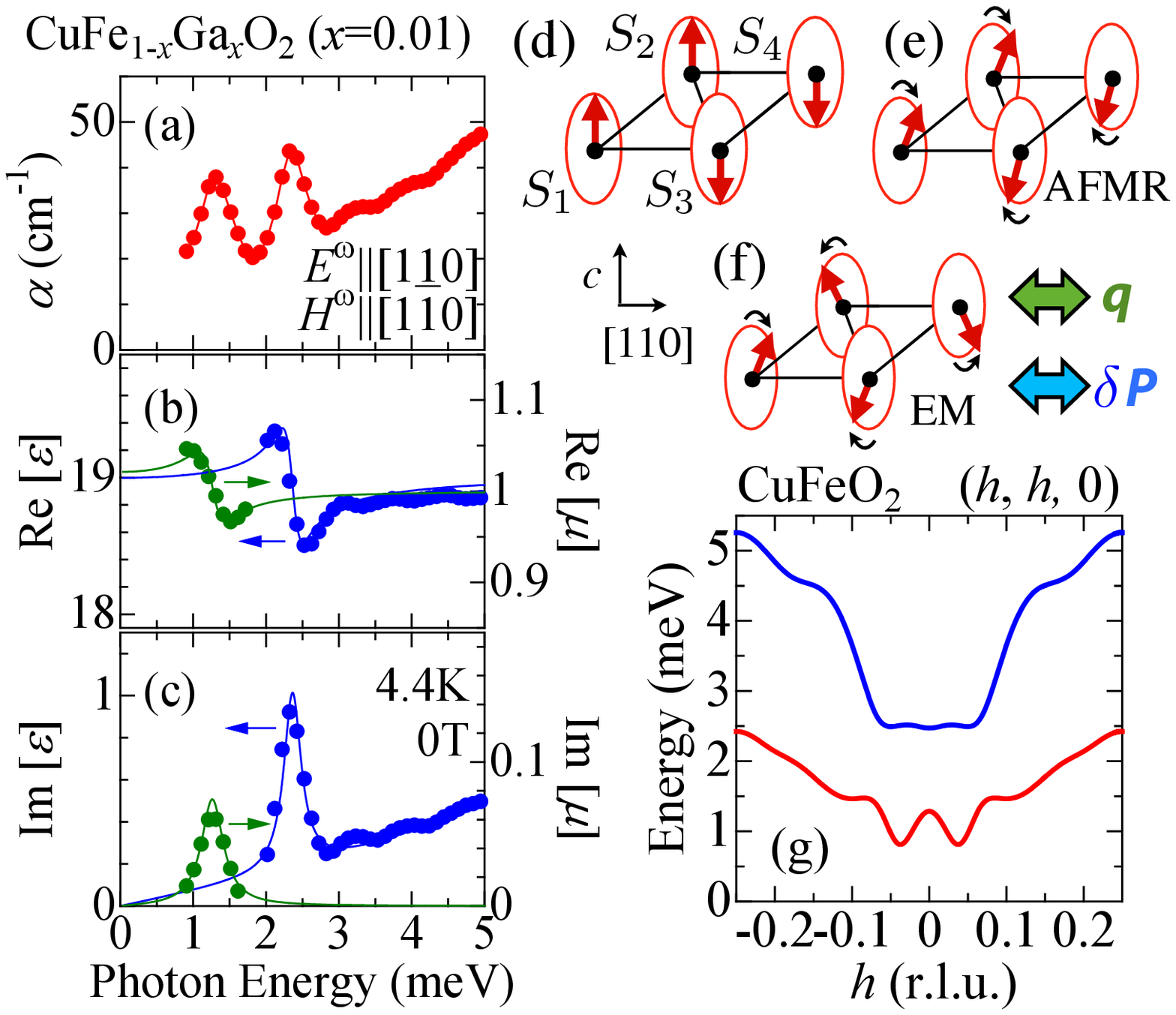}
\caption{(color online). (a)-(c) Absorption coefficient $\alpha$, real and imaginary part of $\epsilon$ and $\mu$ spectra for the $x = 0.01$ specimen at 4.4 K with $\Ew \parallel [110]$ and $\Hw \parallel [1\bar{1}0]$. Solid lines in (b) and (c) represent the fits with the sum of Lorentzian functions. (d) Spin structure of the CM4 magnetic ground state. (e) and (f) indicate the possible excitation modes corresponding to the observed genuine magnon(AFMR) and electromagnon(EM), respectively. In (f), $S_1$ and $S_3$ rotate to the opposite direction of $S_2$ and $S_4$ within a plane perpendicular to $\vec{q}$. (g) Spin-wave (magnon) dispersion of {\CFO} along the ($h$, $h$, 0) direction as suggested in Ref. \cite{CuFeO2_SpinWave2}. }
\end{center}
\end{figure}

\begin{figure}
\begin{center}
\includegraphics*[width=8cm]{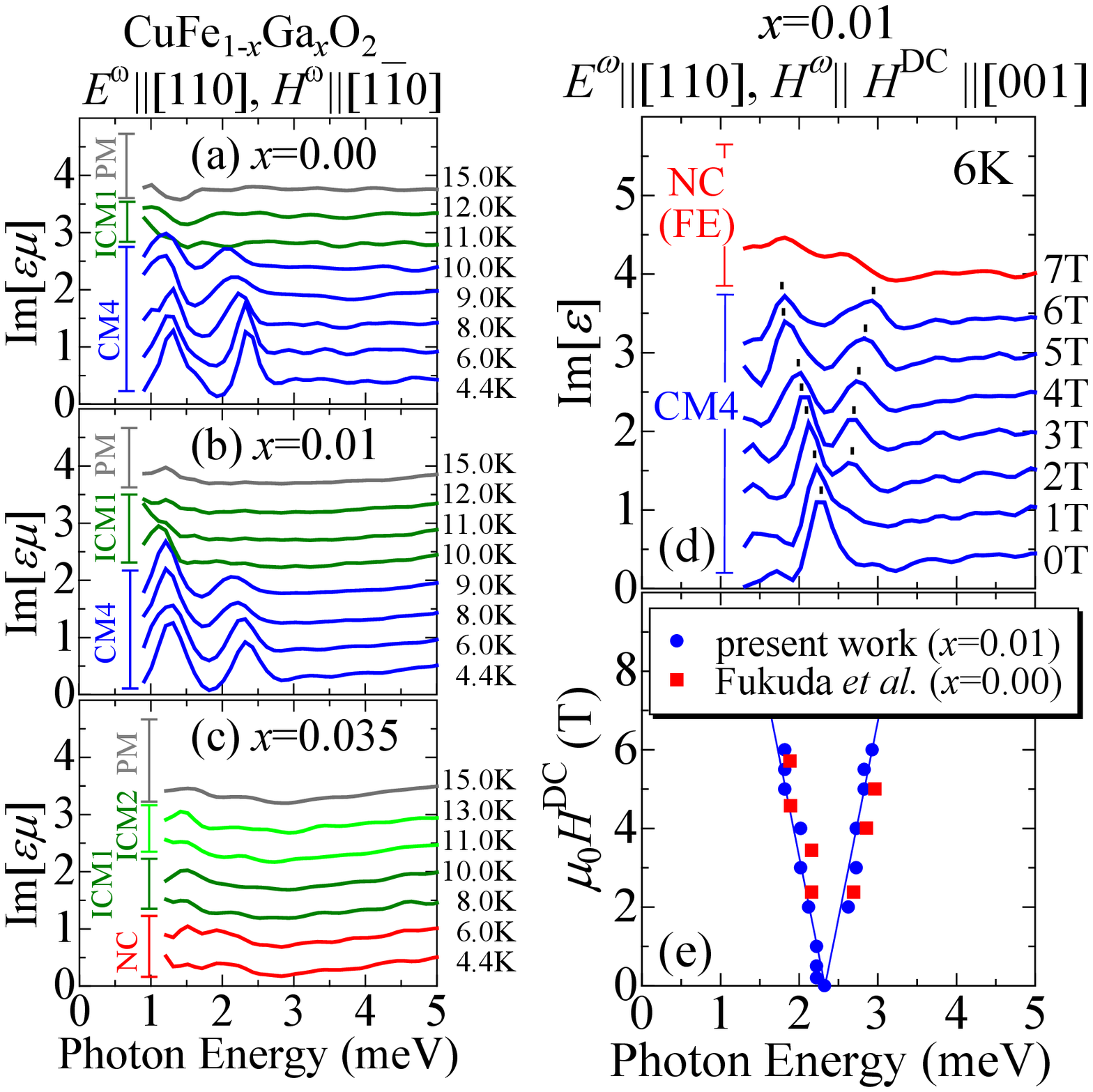}
\caption{(color online). $T$-dependence of Im[$\epsilon \mu$] spectra measured with $\Ew \parallel [110]$ and $\Hw \parallel [1\bar{1}0]$ for (a) $x=0.00$, (b) $x=0.01$, and (c) $x=0.035$ specimens, respectively. (d) Im[$\epsilon$] spectra for the $x=0.01$ specimen measured at 6 K with $\Ew \parallel [110]$ and $\Hw \parallel [001]$ in various magnitudes of static $H$ ($H^\textrm{DC}$) applied along the [001] direction. $H$-dependence of observed EM peak positions (circle), as well as the development of resonance modes previously reported by the ESR study for the $x = 0.00$ specimen \cite{CuFeO2_ESR} (square), are plotted in (e).} 
\end{center}
\end{figure}

Next, we discuss the microscopic origin of these excitations. The spin-wave (SW) dispersion for CuFeO$_2$ has been investigated by a previous inelastic neutron diffraction study\cite{CuFeO2_SpinWave1}, and the analysis clarified the existence of two SW branches as reproduced in Fig. 3(g)\cite{CuFeO2_SpinWave2}. In general, an antiferromagnetic resonance (AFMR) appears as the excitation of zone center mode at $\vec{k}=0$ by $\Hw$ perpendicular to the collinear spin direction(Fig. 3(e)). From this criterion, we concluded the excitation at 1.2 meV driven by $\Hw \parallel [1\bar{1}0]$ is AFMR on the lower SW branch. In contrast, the excitation energy of the observed EM ($\sim 2.3$ meV) agrees with that of the zone-center mode on the upper SW branch.

So far, the most successful scheme to explain the dynamical ME coupling is the exchange striction. However, this mechanism is inactive in  the collinear spin system like the present CM4 phase, since the relationship $\delta P_{ij} \propto S_i \cdot \delta S_j = 0$ always holds\cite{EM_Striction, EM_Mochizuki}. This strongly suggests the relevance of spin-orbit coupling to the present EM mode; the relatively weak peak intensity in the Im[$\epsilon$] spectrum (one order of magnitude smaller than that of DyMnO$_3$\cite{EM_DyMnO3}) also supports this scenario. Since the static $P$ in NC(FE) is induced by the proper screw magnetic order through the spin-orbit-interaction mediated modulation of Fe $3d$ - O $2p$ hybridization\cite{Jia1, CuFeO2_Arima}, we may anticipate the analogous origin for the presently observed dynamical ME coupling in CM4. For example, the magnetic excitation as depicted in Fig. 3(f) can dynamically generate a proper-screw-like spin texture with a finite spin chirality, which is expected to induce non-zero electric dipole along the $\delta P \parallel q \parallel [110]$ direction. This mode should be active only with $E^\omega \parallel [110]$, which is consistent with the experimental results. The disappearance of EM mode in the NC(FE) phase may reflect the alteration of magnetic symmetry or Brillouin-zone folding, but the detail is left to be clarified.

We further investigated the development of the EM mode in static $H$ applied along the [001] direction. Figure 4(d) indicates the $H$-dependence of Im[$\epsilon$] spectrum measured at 6 K for the $x=0.01$ specimen with $\Ew \parallel [110]$ and $\Hw \parallel [001]$, where only the EM excitation can be observed. As $H$ increases, the EM mode is found to split linearly with $H$ and form two peak structures. This reflects the $H$-linear splitting of SW branches, which is generally expected for collinear antiferromagnets with $H$ applied parallel to the magnetic easy axis. Since the spectral shape of the two-magnon excitation should be independent of $H$\cite{TwoMagnon}, this ensures that the present EM mode is excited by the one-magnon process. The observed evolution of EM peak positions under applied $H$ is summarized in Fig. 4(e). Note that similar $H$-dependence of resonance modes has been reported by Fukuda {\etal} from the ESR study for the $x = 0.00$ specimen\cite{CuFeO2_ESR}, while they conventionally assigned these modes to AFMR driven by $\Hw$. Our present results imply that the resonance modes found in the previous ESR study\cite{CuFeO2_ESR} is primarily driven by $\Ew$-component of incident microwave. The peak structure observed in Im[$\epsilon$] spectrum becomes almost invisible after the transition from CM4 into NC(FE) at 6.3 T, which confirms the inactivity of EM mode in the latter NC phase.


In conclusion, we have experimentally revealed the electromagnon excitation in the paraelectric $\uparrow \uparrow \downarrow \downarrow$ collinear magnetic phase of triangular lattice antiferromagnet {\CFGO}. This mode was found to vanish in the ferroelectric helimagnetic phase. These facts prove that neither ferroelectricity nor noncollinear magnetism is a necessary condition for the appearance of electromagnon excitation, while the existing theories on electromagnon have focused on noncollinear magnets like helimagnets. The anti-correlation between the noncollinear magnetism and the emergence of electromagnon excludes the exchange striction mechanism as the origin of dynamical ME coupling: The electric activity of the magnon in this compound is ascribed to the modulation of the $p$-$d$ hybridization at the spin-twisted excited state via the spin-orbit interaction. Our discovery suggests that similar electromagnon modes will be observable in a wide range of paraelectric collinear magnets.

\section{Appendix}

In this appendix, we provide the detail of the terahertz time-domain spectroscopy and discuss the validity of our transformation procedure among different optical constants.

In the present study, terahertz time-domain spectroscopy (THz-TDS) was performed in the transmission geometry. The detailed experimental setup is described in Ref. \cite{EM_DyMnO3}. We employed the photo-conducting antennas made of low-temperature-grown GaAs as emitters and detectors, to access the energy down to 1 meV. The direction of the light-polarization was carefully set parallel to the crystallographic axis using a wire grid polarizer.

\begin{figure}[h]
\begin{center}
\includegraphics*[width=6cm]{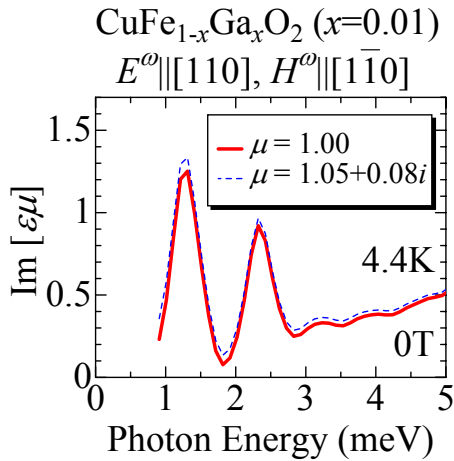}
\caption{(color online). Imaginary part of $\epsilon \mu$ spectra for the $x = 0.01$ specimen measured at 4.4 K with $\Ew \parallel [110]$ and $\Hw \parallel [1\bar{1}0]$ in zero magnetic field. For a solid (dashed) line, $\mu = 1.0$ ($\mu = 1.05 + 0.08 i$) is assumed for the pre-exponential factor in Eq. (2), respectively. Note that the solid line represents the same data as shown in Fig. 2(b).}
\end{center}
\end{figure}

In THz-TDS, wave form of irradiated pulse electric field (ranging within a few picoseconds) is directly measured in time domain with and without the specimen. They are converted into the frequency domain via the fast Fourier transformation (FFT), and the spectra of complex transmission constant ($t$) is deduced using the relationship $t = E^{\omega}_{\textrm{sample}} / E^{\omega}_{\textrm{ref}}$. Here, we can avoid the effect of multiple reflections and resultant interference by restricting the time range of FFT. Obtained $t$ is further transformed into complex refractive index $n=\sqrt{\epsilon \mu}$ using Eq. (2). Since Eq. (2) is not explicitly solvable, we made its pre-exponential factor approximate $4n / (n+1)^2$ by assuming $\mu \simeq 1$ unless otherwise noted.

This approximation hardly affects the obtained $n$ (or $\epsilon \mu$) spectrum, as exemplified in the following by the spectra for the $x=0.01$ specimen with $\Ew \parallel [110]$ and $\Hw \parallel [1\bar{1}0]$ at 4.4 K. A solid line in Fig. 5 indicates the imaginary part of $\epsilon \mu$-spectrum deduced with the aforementioned  $\mu \simeq 1$ approximation. Its decomposed $\epsilon$- and $\mu$-spectra, which are derived with the method described in the main text, are also shown in Figs. 3 (b) and (c). We find that the relationship $|\mu -1| < 0.1$ always holds, which justifies the presently assumed $\mu \simeq 1$ condition. To further check the validity of this approximation, we deduced $\epsilon \mu$ spectrum assuming $\mu = 1.05 + 0.08i$ (the maximum $\mu$-value taken from Figs. 3 (b) and (c)) for the pre-exponential factor in Eq. (2). The result is plotted in Fig. 5 as a dashed line, which is consistent with the one calculated with the original $\mu = 1$ assumption (the solid line).

\section{Acknowledgements}

The authors thank T. Arima, S. Miyahara, Y. Takahashi, Y. Kohara, and M. Uchida for enlightening discussions. This work was partly supported by Grants-In-Aid for Scientific Research (Grant No. 20340086, 2010458) from the MEXT of Japan, and FIRST Program by the Japan Society for the Promotion of Science (JSPS) .

\end{document}